\documentstyle[prd,aps,floats,epsf]{revtex}

\def\bea{\begin{eqnarray}}
\def\eea{\end{eqnarray}}

\def\tQ{\tilde Q}

\begin{document}
\preprint{BROWN-HET-1142, hep-ph/yymmddd}
\draft 

%
%
 
%
\renewcommand{\topfraction}{0.99}
\renewcommand{\bottomfraction}{0.99}
\twocolumn[\hsize\textwidth\columnwidth\hsize\csname 
@twocolumnfalse\endcsname
  
\title
{\Large Parametric Amplification of Gravitational Fluctuations During Reheating}
       
\author{F. Finelli$^{1,2}$ and R. Brandenberger$^1$} 
\address{~\\$^1$Department of Physics, Brown University, 
Providence, RI 02912, USA;
~\\$^2$Department of Physics and INFN, Bologna, ITALY}
\date{\today} 
\maketitle
\begin{abstract} 
We demonstrate that cosmological perturbations can undergo amplification by parametric resonance during the preheating period following inflation, even on scales larger than the Hubble radius, without violating causality. A unified description of gravitational and matter fluctuations is crucial in order to determine the strength of the instability. To extract specific signatures of the oscillating inflaton field during reheating, it is essential to focus on a variable describing metric fluctuations which is constant in the standard analyses of inflation. For a massive inflaton without self coupling, we find no additional growth of super-horizon modes during reheating beyond what the usual analyses of the growth of fluctuations predict. However, for a massless self coupled inflaton, there is an initial increase in the value of the gravitational potential, which may lead to different predictions for observations.        
\end{abstract}

]

\vskip 0.4cm
{\bf 1.} As was initially realized in \cite{TB} and worked out in detail in
\cite{KLS94,STB,KLS97} and many other papers \cite{others}, parametric resonance instabilities play a crucial role in the dynamics of the reheating of an inflationary Universe. Reheating (or now more accurately called ``preheating" \cite{KLS94}) is the period after inflation when the inflaton, the scalar field driving the period of exponential expansion, oscillates coherently about its ground state and gradually transforms its energy into the matter and radiation which the present Universe is made up of.

The parametric resonance instability can be seen by considering the linearized equation of motion of fields $\chi$ which couple to the inflaton. Neglecting for a moment the expansion of the Universe, the equation of motion for the Fourier modes $\chi_k$ becomes a harmonic oscillator equation with a periodically varying mass, the Mathieu equation. It is well known that this equation admits instability bands, regions of $k$ for which the solutions grow exponentially. As already noted in \cite{TB,KLS94,STB} and discussed in detail in \cite{KLS97}, the instabilities persist after taking the expansion of the Universe into account.

The parametric resonance instabilities have important consequences for cosmology. They will lead to a reheating temperature which can be much larger than would be obtained by calculating the efficiency of reheating using perturbation theory, as was done initially in \cite{ABF,DL}. This could have important implications for GUT-scale baryogenesis \cite{KLR}. Since parametric resonance only amplifies the amplitude of certain frequency bands of excitations, the state produced will initially have a nonthermal frequency distribution. This may lead to the possibility of the formation of topological defects after inflation \cite{KLS96,Tkachev,KKLT,TKKL,PS98,KK97}, and to the production of supermassive dark matter \cite{CKR}.

Initially \cite{TB}, the parametric instability was discussed in a toy model of new inflation in which the resonance bands were narrow. It was then pointed out \cite{KLS94} that in more realistic models of chaotic inflation, the instability bands were in fact much wider (``broad resonance"), and the Floquet exponent $\mu_k$ giving the rate of exponential growth correspondingly larger. In \cite{PR} it was discovered that a negative sign of the coupling constant between the inflaton and the $\chi$ field leads to a greatly enhanced instability (``negative coupling instability"). White noise eliminates the stability bands alltogether \cite{ZMCB,Bassett1} and mode by mode strengthens the resonance.

The inflaton field also couples to linearized metric perturbations. Bassett \cite{Bassett2} first realized that gravitational waves also experience parametric resonant amplification during reheating. Scalar metric perturbations, however, constitute the components of the metric which couple directly to the matter sector. Hence, it is to be expected that scalar perturbations are amplified more strongly than gravitational waves. The growth of scalar perturbations due to the oscillations of the inflaton field was first considered by Nambu and Taruya \cite{NT96} and by Kodama and Hamazaki \cite{KH96} (see also \cite{KHb96}). Nambu and Taruya concluded that scalar perturbations are amplified during reheating, but did not compare their growth with the usual growth of cosmological perturbations. Kodama and Hamazaki focused on the evolution of the ``Bardeen parameter", a gauge invariant measure of the cosmological perturbations which in the usual analysis of the growth of fluctuations (which neglects the oscillations of the inflaton field) is constant in time for modes with wavelength larger than the Hubble radius. They concluded that in spite of the singular behavior of the the quantity $c_s^2 = {\dot p} / {\dot \rho}$, where $p$ and $\rho$ are pressure and energy density, respectively, the Bardeen parameter remains constant. On the other hand, in a very interesting recent letter, Bassett, Kaiser and Maartens \cite{BKM} have re-analyzed this problem and argue that there is a negative coupling parametric resonance instability which leads to a rapid growth of metric perturbations which in turn act as a pump for matter perturbations. Consequently, matter perturbations are amplified parametrically even in matter models in which there is no resonance in the absence of gravitational fluctuations.

In this letter, we analyze the growth of metric inhomogeneities during reheating in a more complete way, making use of the gauge-invariant theory of perturbations (see \cite{MFB} for a review).  Since matter and metric are coupled by the Einstein constraint equations, the fluctuations can be described completely by a single gauge-invariant variable $\Phi$. In longitudinal gauge, the perturbed metric can be written in terms of $\Phi$ as
\begin{equation} \label{metric}
ds^2 \, = \, dt^2 (1 + 2 \Phi) - a^2(t)(1 - 2 \Phi) \, ,
\end{equation}
where $a(t)$ is the scale factor. The equation of motion for $\Phi$ (linearized Einstein-Higgs equation) is a linear second order differential equation. As pointed out in \cite{KH96}, two of the coefficients in this equation of motion are singular due to the oscillations of the inflaton field. This is a problem encountered a long time ago \cite{axion} in the context of axion fluctuations. As realized in \cite{NT96,KH96}, the divergence disappears if, instead of $\Phi$, one considers the equation of motion for the Mukhanov variable $Q$ \cite{Mukhanov}, a variable in terms of which the quantization of cosmological perturbations is straightforward (see \cite{MFB} for a review).
We demonstrate that an instability persists in the equation of motion for a rescaled variable ${\tQ}$. This instability, however, is not of negative coupling type. For a massive inflaton, it only leads to an increase of ${\tQ}$ proportional to $a^{3/2}(t)$ for long wavelength fluctuations. Hence, the amplitude of $Q$ is constant in time, and there is no amplification of fluctuations beyond what the usual theory (which does not take the details of reheating into account) predicts. However, for a massless self coupled inflaton, ${\tQ}$ experiences an initial increase. From the regular behavior of $Q$, a completely regular behavior of $\Phi$ can be deduced. We show why the usual methods to study the evolution of perturbations in inflationary cosmology miss additional growth of fluctuations due to the oscillating inflaton field.
        
{\bf 2.} Our starting point are the equations of motion for the perturbations of the Einstein-Higgs system about a Friedmann-Robertson-Walker background solution. In terms of the gauge invariant metric and matter variables $\Phi$ (see (\ref{metric})) and $\delta \phi_{\rm{gi}}$ (which in longitudinal gauge is equal to the scalar field perturbation $\delta \phi$, because of which we will subsequently drop the subscript), the system of equations in momentum space is
\begin{equation}  
{\ddot \Phi} + 3H{\dot \Phi} + \bigl[{{k^2} \over {a^2}} + 2({\dot H} + H^2)\bigr]
\Phi \, = \, \kappa^2 ({\ddot \phi} + H{\dot \phi})\delta\phi \label{e2}
\end{equation}
\begin{equation}
{\ddot{\delta\phi}} + 3H{\dot{\delta\phi}} + ({{k^2} \over {a^2}} + V'')\delta\phi \,= \, 4{\dot \Phi}{\dot \phi} - 2V'\Phi \label{e3}
\end{equation}
\begin{equation}
{\dot \Phi} + H\Phi \, = \, {1 \over 2}\kappa^2{\dot \phi}{\delta\phi} \, , \label{e4}
\end{equation}
where $\kappa^2 = 8{\pi}G$, $H = {\dot a}/a$ is the Hubble expansion rate, $\phi$ is the homogeneous background field for a scalar matter field with potential energy density $V(\phi)$, and where a prime denotes the derivative with respect to $\phi$. Equation (\ref{e3}) is the equation of motion for $\delta\phi$ in a perturbed metric, (\ref{e4}) is the Einstein momentum constraint equation, and (\ref{e2}) is a combination of the dynamical equation of motion for $\Phi$ and the Einstein energy constraint equation (see eqs. (6.42) and (6.40) of \cite{MFB}). These are exactly the same equations (except expressed in terms of physical time) as eqs. (2 - 4) of \cite{BKM}.

Because of the Einstein constraint equation (\ref{e4}), there is only one physical degree of freedom which we can choose to be $\Phi$. Note that the source term in (\ref{e2}) is not suppressed compared to the terms on the left hand side of the equation in spite of the factor of $\kappa^2$ which multiplies the term, since the constraint equation involves a compensating factor of $\kappa^2$. The correct equation of motion for $\Phi$ is obtained by inserting the constraint equation (\ref{e4}) into (\ref{e2}), with the result
\begin{equation} \label{e5}
{\ddot \Phi} + (H - 2{{\ddot \phi} \over {\dot \phi}}){\dot \Phi} +
({{k^2} \over {a^2}} + 2{\dot H} - 2H{{\ddot \phi} \over {\dot \phi}})\Phi \, = \, 0 \, .
\end{equation}

During the slow rolling period of an inflationary cosmology, the coefficients in this equation are well-behaved. However, oscillations of $\phi$ during reheating lead to singularities. As was realized in \cite{NT96,KH96}, these singularities can be eliminated by making use of Mukhanov's variable \cite{Mukhanov} $Q$, the combination 
\begin{equation} \label{e6}
Q \, = \, \delta\phi + {{\dot \phi} \over H}\Phi
\end{equation}
of the gauge invariant matter and metric perturbations in terms of which the unified quantization of the matter and metric perturbations is easy. In terms of $Q$, the equation of motion (\ref{e5}) becomes \cite{MFB,NT96,KH96}
\begin{equation} \label{e7}
{\ddot Q} + 3H{\dot Q} + \bigl(V'' + {{k^2} \over {a^2}} + 2({{\dot H} \over H} + 3H)^{.}\bigr)Q \, = \, 0 \, .
\end{equation}
As is evident, the coefficients of this differential equation are regular. Given $Q$, it is possible to obtain $\Phi$ since (\ref{e6}) can be rewritten in the form
\begin{equation} \label{e8}
{{k^2} \over {a^2}}\Phi \, = \, {{\kappa^2} \over 2}{{{\dot \phi}^2} \over H}\bigl({H \over {\dot \phi}}Q\bigr)^{.} \, .
\end{equation}

The Hubble damping term in the equation (\ref{e7}) for $Q$ can be eliminated by introducing the rescaled variable ${\tQ}$. For a massive inflaton with potential $V(\phi) = m^2 \phi^2 / 2$, $\tQ = a^{3/2} Q$.  In terms of $\tQ$, (\ref{e7}) becomes
\begin{equation} \label{rescaledv}
{\ddot{\tQ}} + \bigl[V'' + {{k^2} \over {a^2}} + 2({{\dot H} \over {H}} + 
3H)^{.} - {9 \over 4}\bigl(H^2 + 
{2 \over 3}{\dot H}\bigr)\bigr]\tQ \, = \, 0 \, .
\end{equation}
Making use of the background Einstein equations, (\ref{rescaledv}) can be written as
\begin{eqnarray} \label{rescaledv2}
{\ddot{\tQ}} &+& \bigl[V'' + {{k^2} \over {a^2}} + 3\kappa^2{\dot \phi}^2 \\
&+& 2\kappa^2{{{\dot \phi}V'} \over H} - {{\kappa^4} \over {2H^2}}{\dot \phi}^4 
+ {{3 \kappa^2} \over 4}{p_{\phi}}\bigr]{\tQ} \, = \, 0 \, , \nonumber
\end{eqnarray}
where $p_{\phi}$ is the background pressure of the scalar field.

After the period of slow rolling has ended, the value of $H$ is smaller than $m$. Hence, it follows from the background equation of motion for $\phi$ that - in the absence of back-reaction and with accuracy increasing in time - the motion of ${\tilde \phi} = a^{3/2}\phi$ is oscillatory in time. In this limit,
equation (\ref{rescaledv}) has the form 
\begin{equation} \label{massive}
{\ddot{\tQ}} + \bigl[A(k) - 2q{\rm cos(mt)}\bigr]{\tQ} \, = 0 \, ,
\end{equation}
where
\begin{equation}
A(k) \, = \, m^2 + {{k^2} \over {a^2}} + r \, ,
\end{equation}
where $r$ contains the time average of the last four terms in the square bracket of (\ref{rescaledv2}), and $q$ contains the coefficients of the oscillating parts of these terms. Since $q(t)$ is decreasing, (\ref{massive}) is not of the form of the usual parametric resonance equation, and no exponentially growing solutions will result. 

The second of the four last terms in the square bracket of (\ref{rescaledv2}) is the most important. Its initial amplitude is the largest, and it decays the least fast as a function of time. Approximating $a(t) \sim t^{2/3}$ corresponding to a pressureless phase, it is easy to check that the second term decays as $t^{-1}$, whereas the other three terms decay as $t^{-2}$. Note that the decay rate of the dominant term of $q$ as a function of time is less fast than the corresponding decay rate of $q$ for matter fluctuations \cite{KLS97}, a point already emphasized by \cite{BKM}. The amplitude of this leading term in $q$ starts out slightly larger than $m^2$. Due to the expansion of the Universe, the instability does not lead to exponential increase in ${\tQ}$, but only to an increase proportional to $a^{3/2}(t)$ (see Figure 1), which implies that the amplitude of $Q$ remains constant. 
 
The results are different for a massless self coupled inflaton with potential $V(\phi) = \lambda \phi^4 / 4$. In this case, we eliminate (following \cite{GKLS}) the Hubble damping term in (\ref{e7}) by introducing the rescaled variable ${\tQ} = a(\eta)Q$ and by working in terms of conformal time $\eta$. In this case, (\ref{e7}) becomes
\begin{equation} \label{massless}
{\ddot {\tQ}} + \bigl[A(k) + 3 cn^2(x,{1 \over {\sqrt 2}})\bigr]\tQ \, = 0 \, ,
\end{equation}
where $x = \sqrt{\lambda}{\cal A}_{\phi}\eta$ is a rescaled conformal time, ${\cal A}_{\phi}$ is the amplitude of oscillation of $a(\eta)\phi$, $cn$ stands for the elliptic cosine function, and
\begin{equation}
A(k) \, = \, {{k^2} \over {\lambda {{\cal A}_\phi}^2}} + s \, .
\end{equation}
Here, $s$ stands for oscillatory terms which are important initially and lead to an increase in $Q$ (even for $k^2 \ll \lambda {\cal A}_{phi}^2$), but which decay in time as $\eta^{-1}$. For $s = 0$, Equation (\ref{massless}) is a Lam\'e equation, has been studied in detail in \cite{GKLS} and exhibits no resonance for $k^2 \ll \lambda {\cal A}_{phi}^2$.  
 
{\bf 3.}  We have solved the equation of motion (\ref{rescaledv2})
numerically. Figure 1 shows the resulting time evolution of the rescaled variable ${\tQ}$ in the case of a massive inflaton for the mode $k = 0.1 m$ (about six times larger wavelength than the Hubble radius) over a period of several oscillations of the background field. As mentioned above, the amplitude of ${\tQ}$ grows linearly in time (and thus the amplitude of $Q$ remains constant) to within the numerical accuracy.   

\begin{figure}
\epsfxsize=2.9 in \epsfbox{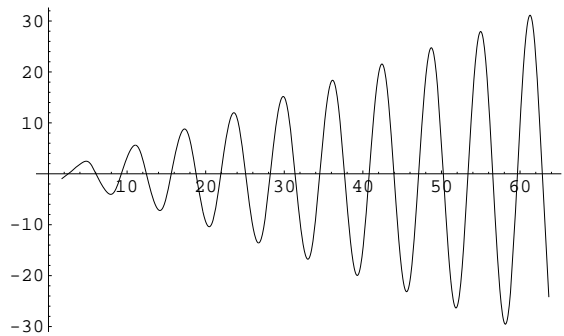}
\caption{Evolution of $\tQ$ as a function of time in the case of a massive inflaton for a mode with $k = 0.1 m$,
with initial conditions such that $\tQ$ is $-1$ and its time derivative 0 at the initial time when $\phi = 0.2 m_{pl}$, after slow rolling has ended. Time is expressed in units of $m^{-1}$.}
\label{fig1}
\end{figure}

Does the growth of ${\tQ}$ imply that there is an increase in the amplitude of cosmological fluctuations beyond what is calculated in the usual analyses which model the reheating as a continuous change in the equation of state without any oscillations? According to the usual treatment (see e.g. \cite{MFB}), the details of the equation of state are completely unimportant for the final amplitude of fluctuations on scales larger than the Hubble radius during the reheating period. If there were an increase in the amplitude of $\Phi$ beyond what is usually predicted, this could have important implications for the spectrum of density perturbations and microwave anisotropies, as conjectured in \cite{BKM}. First of all, let us note that such an effect would NOT violate causality. Inflation has already set up fluctuations on all scales smaller than the Hubble radius at the beginning of inflation, and in modelling the Universe as a Friedmann Universe, we have inserted correlations on even larger scales. These perturbations can self gravitate and increase in magnitude during reheating without violating causality. This is precisely the way in which quantum fluctuations created during inflation become large after inflation. Since the phase of oscillation of $\phi$ is coherent over a region much larger than the Hubble radius during reheating, it is therefore possible that the oscillations might have an effect on modes larger than the Hubble radius.

The clearest way to determine if the growth of ${\tQ}$ calculated above is a new effect is to calculate the quantity $\zeta$ 
\begin{equation} \label{zeta}
\zeta \, = {2 \over 3}{{\Phi + H^{-1}{\dot \Phi}} \over {1 + w}} + \Phi \, , 
\end{equation}
which according to the standard theory of the growth of cosmological perturbations remains constant for modes with wavelength larger than the Hubble radius \cite{zetaref}. Here, $w = p / \rho$ describes the equation of state, and $p$ and $\rho$ are the pressure and energy density, respectively. More precisely \cite{MFB}, ${\dot \phi}^2{\dot \zeta} = 0$ is equivalent to the equation of motion for $\Phi$ (for modes with wavelength larger than the Hubble radius). Hence, it is usually deduced that ${\dot \zeta} = 0$. This conclusion, however, may break down if ${\dot \phi} = 0$ which occurs precisely in the phase we are studying here during which $\phi$ is oscillating coherently. Hence, it is possible that additional resonant amplification of fluctuations during reheating occurs.

The variable $\zeta$ is related to $Q$ via
\begin{equation} \label{zetaeq}
\zeta \, = \, {H \over {\dot \phi}}Q \, .
\end{equation}
Since the amplitudes of $H$ and ${\dot \phi}$ decrease at the same rate (proportional to $t^{-1}$) if we assume that $a(t) \sim t^{2/3}$, the constancy of the amplitude of $Q$ (for a massive inflaton) leads to the conclusion that $\zeta$ is constant when evaluated at the same phase during each oscillation period of the inflaton $\phi$. Since $1 + w$ has constant amplitude during the period of oscillation, it follows from the constancy of $\zeta$ that the amplitude of $\Phi$ is constant, inasfar as the ${\dot \Phi}$ term in (\ref{zeta}) can be neglected. This can also be seen by evaluating $\Phi$ directly using (\ref{e8}) (see \cite{FPEB}) for further discussion). 

Note that $\Phi$ is the basic physical quantity which is well-defined at all times. It is the quantity which determined the power spectrum of density fluctuations and of CMB anisotropies. In contrast, $\zeta$ is an auxiliary quantity. At each zero of ${\dot \phi}$ there is a singularity in the relation between $\Phi$ and $\zeta$. What is therefore important is to calculate the value of $\zeta$ for each zero crossing of $\phi$ and compare the values. In Figure 2 we plot $(1 + w)\zeta$ as a function of time, determined directly from (\ref{zetaeq}). The fact that the amplitude of oscillation of this function is constant implies that $\zeta$ does not change over a period. This demonstrates that the growth of ${\tQ}$ observed in Figure 1 is in exact agreement with the usual analysis of the growth of cosmological perturbations.  

\begin{figure}
\epsfxsize=2.9 in \epsfbox{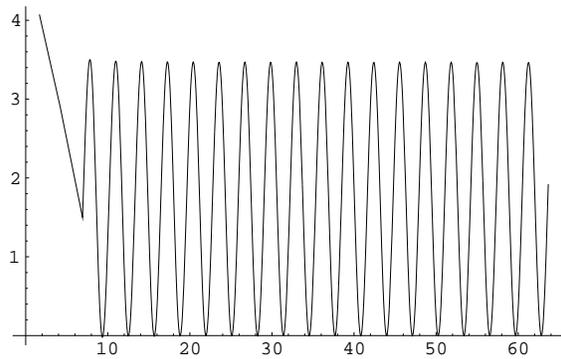}
\caption{Evolution of $(1 + w)\zeta$ for a theory with a massive inflaton, and for same mode, initial conditions and initial time as in Fig. 1.}
\label{fig2}
\end{figure}
 
{\bf 4.} To summarize, we have studied the growth of cosmological perturbations during a phase of coherent oscillations of the scalar field driving inflation, which leads to singularities in $c_s^2 = {\dot p} / {\dot \rho}$. 
In this period, the equations of motion of the fluctuations are similar in structure to the equations describing coupled matter fields during reheating. As emphasized in \cite{KLS97,PR,GKLS}, the presence or absence of a mass in the inflaton field is crucial. To study the evolution of cosmological perturbations, it is crucial to work in terms of variables in terms of which the singularities are absent, and in which the usual growth of cosmological fluctuations (obtained without taking into account the oscillations of $\phi$) is factored out, such as $\zeta$.

For a massive inflaton, we find no amplification of long wavelength fluctuations (wavelength larger than the Hubble radius during reheating) - the amplitude of $\zeta$ is constant. A modelling of the reheating period including the oscillations of $\phi$ will lead to the same growth as is obtained in the usual analyses of perturbations in which the transition from the inflationary phase to the post-inflationary radiation-dominated period is modelled (implicitly) by a monotonous transition in the equation of state. In a companion letter \cite{EP98}, Parry and Easther demonstrate that even a full nonlinear analysis does not lead to any additional growth of fluctuations with wavelength larger than the Hubble radius during reheating. 

However, for a massless self coupled inflaton, initially both the gravitational potential $\Phi$ and $\zeta$ grow for modes larger than the Hubble radius during reheating, as conjectured in \cite{BKM} (see \cite{FPEB} for a detailed discussion).
 
Note that the negative coupling instability discussed in \cite{BKM} is not present in the relevant equation (\ref{e7}) for ${\tQ}$. The negative coupling instability in the equation for $\Phi$ is precisely the instability which is responsible for the amplification of the fluctuations in the standard analysis of the growth of cosmological perturbations.    

We are grateful to Matthew Parry, Richard Easther, David Kaiser and Bill Unruh  for stimulating discussions, and wish to thank Matthew Parry for independently  verifying many of the equations. One of us (F.F.) also wishes to thank Gianpaolo Vacca for useful suggestions. This work was supported in part (at Brown) by the U.S. Department of Energy under Contract DE-FG02-91ER40688, TASK A, and by 
{\it Borsa di Perfezionamento} from the University of Bologna.


\begin{thebibliography}{10}

\bibitem{TB} J. Traschen and R. Brandenberger, {\it Phys. Rev.} {\bf D42},
2491 (1990).
 
\bibitem{KLS94} L. Kofman, A. Linde and A. Starobinsky, {\it Phys. Rev. Lett.} 
{\bf 73}, 3195 (1994), hep-th/9405187. 

\bibitem{STB} Y. Shtanov, J. Traschen and R. Brandenberger, 
{\it Phys. Rev.} {\bf D51}, 5438 (1995), hep-ph/9407247.
 
\bibitem{KLS97} L. Kofman, A. Linde and A. Starobinsky, 
{\it Phys. Rev.} {\bf D56}, 3258 (1997), hep-ph/9704452 (1997). 

\bibitem{others} M. Yoshimura, {\it Prog. Theor. Phys.} {\bf 94}, 873 (1995);\\
D. Boyanovsky, H. de Vega and R. Holman, ``Erice Lectures on Inflationary Reheating", hep-ph/9701304, and refs. therein;\\
D. Kaiser, {\it Phys. Rev.} {\bf D53}, 1776 (1996).

\bibitem{ABF} L. Abbott, E. Farhi and M. Wise, {\it Phys. Lett.} {\bf 117B}, 29 (1982).  

\bibitem{DL} A. Dolgov and A. Linde, {\it Phys. Lett.} {\bf 116B}, 329 (1982).

\bibitem{KLR} E. Kolb, A. Linde and A. Riotto, {\it Phys. Rev. Lett.} {\bf 77}, 4290 (1996).    

\bibitem{KLS96} L. Kofman, A. Linde and A. Starobinsky, {\it Phys. Rev. Lett.} {\bf 76}, 1011 (1996).     

\bibitem{Tkachev} I. Tkachev, {\it Phys. Lett.} {\bf B376}, 35 (1996).

\bibitem{KKLT} S. Khlebnikov, L. Kofman, A. Linde and I. Tkachev, {\it Phys. Rev. Lett.} {\bf 81}, 2012 (1998).  

\bibitem{TKKL} I. Tkachev, S. Khlebnikov, L. Kofman and A. Linde, hep-ph/9805209.

\bibitem{PS98} M. Parry and A. Sornborger, hep-ph/9805211.

\bibitem{KK97} S. Kasuya and M. Kawasaki, {\it Phys. Rev.} {\bf D56}, 7597 (1997), and also hep-ph/9804429.

\bibitem{CKR} D. Chung, E. Kolb and A. Riotto, hep-ph/9805473.

\bibitem{PR} T. Prokopec and T. Roos, {\it Phys. Rev.} 
{\bf D55}, 3768 (1997), hep-ph/9610400;\\
B. Greene, T. Prokopec and T. Roos, {\it Phys. Rev.} {\bf D58}, 6484 (1997).

\bibitem{ZMCB} V. Zanchin, A. Maia, W. Craig and R. Brandenberger, {\it Phys. Rev.} {\bf D57}, 4651 (1997).

\bibitem{Bassett1} B. Bassett, {\it Phys. Rev.} {\bf D58}, 021303 (1998).

\bibitem{Bassett2} S. Khlebnikov and I. Tkachev, {\it Phys. Rev.} {\bf D56}, 653 (1997);\\
B. Bassett, {\it Phys. Rev.} {\bf D56}, 3429 (1997).

\bibitem{NT96} Y. Nambu and A. Taruya, {\it Prog. Theor. Phys.} {\bf 97}, 83 (1997).

\bibitem{KH96} H. Kodama and T. Hamazaki, {\it Prog. Theor. Phys.} {\bf 96}, 949 (1996).

\bibitem{KHb96} T. Hamazaki and H. Kodama, {\it Prog. Theor. Phys.} {\bf 96}, 1123 (1996).

\bibitem{BKM} B. Bassett, D. Kaiser and R. Maartens, {\it General relativistic preheating after inflation}, hep-ph/9808404.

\bibitem{MFB} V. Mukhanov, H. Feldman and R. Brandenberger, {\it Phys. Rep.} {\bf 215}, 203 (1992).

\bibitem{axion} M. Sasaki, {\it Prog. Theor. Phys.} {\bf 72}, 1266 (1984);\\
H. Kodama and M. Sasaki, {\it Prog. Theor. Phys. Suppl.} {\bf 78}, 1 (1984);\\
R. Brandenberger, {\it Phys. Rev.} {\bf D32}, 501 (1985).

\bibitem{Mukhanov} V. Mukhanov, {\it JETP} {\bf 67}, 1297 (1988).

\bibitem{GKLS} P. Greene, L. Kofman, A. Linde, and A. Starobinsky, 
{\it Phys. Rev.} {\bf D56}, 6175 (1997).

\bibitem{zetaref} J. Bardeen, P. Steinhardt and M. Turner, {\it Phys. Rev.} {\bf D28}, 679 (1983);\\
R. Brandenberger and R. Kahn, {\it Phys. Rev.} {\bf D29}, 2172 (1984);\\
D. Lyth, {\it Phys. Rev.} {\bf D31}, 1792 (1985).

\bibitem{FPEB} F. Finelli, M. Parry, R. Easther and R. Brandenberger, in preparation.

\bibitem{EP98} M. Parry and R. Easther, hep-ph/9809574.

\end{thebibliography}
\end{document}